\documentclass{PoS}
\usepackage{psfrag,epsfig,graphicx,graphics,xcolor}
\usepackage{wrapfig}
\usepackage{amsmath}

\title{Phenomenology of the polarized cross-sections of the $\rho$ meson leptoproduction at high energy}

\ShortTitle{Rho meson leptoproduction at high energy}

\author{\speaker{Adrien Besse}
\\
        Irfu - SPhN, CEA Saclay, France\\
        E-mail: \email{adrien.besse@cea.fr}}

\author{Lech Szymanowski\\
        National Centre for Nuclear Research (NCBJ), Warsaw, Poland\\
       E-mail: \email{Lech.Szymanowski@fuw.edu.pl}}
				
\author{Samuel Wallon\\
       LPT, Universit\'e Paris-Sud, CNRS, 91405, Orsay, France\\
       UPMC Univ. Paris 06, facult\'e de physique, 4 place Jussieu, 75252 Paris Cedex 05,
France\\
       E-mail: \email{Samuel.Wallon@th.u-psud.fr}}

\newcommand{\kb}{\underline{k}}

\newcommand{\qb}{\bar{q}}
\newcommand{\rb}{\underline{r}}

\newcommand{\nn}{\nonumber}

\abstract{
We present a model for the polarized cross-sections of the hard diffractive leptoproduction of $\rho$ meson in the high energy limit. Our model is based on the light-cone collinear factorization of the virtual photon to $\rho$ meson impact factor when using the impact factor representation of the helicity amplitudes of the $\rho$ meson leptoproduction. This gauge invariant treatment when expressed in impact parameter space, leads to the factorization on one hand of the color dipole scattering amplitude and on the other hand of the distribution amplitudes of the $\rho$ meson up to twist 2 and 3. We show that the results of this approach are in good agreement with HERA data for virtualities above $\sim 5\;$GeV$^2$.}

\FullConference{Photon 2013,\\
		20-24 May 2013\\
		Paris, France }

\begin{document}

\section{Introduction}
In this study, we provide a model for the helicity amplitudes $T_{\lambda_{\rho}\lambda_{\gamma}}$ of the hard exclusive diffractive leptoproduction of $\rho$ meson\footnote{$\lambda_{\gamma}$ and $\lambda_{\rho}$ denote the virtual photon and the $\rho$ meson polarizations.} $$\gamma^*(q,\lambda_{\gamma})\, N(p)\to \rho(p_{\rho},\lambda_{\rho})\,N(p')\,,$$ in the perturbative Regge limit
$W^2=(p+q)^2\gg Q^2\gg \Lambda_{QCD}^2\,,$ and in the limit of vanishing $t$. In this limit, $T_{00}$ and $T_{11}$ are the only remaining helicity amplitudes, other helicity amplitudes are suppressed at least by a factor $\sqrt{-t}/Q$.  On the experimental side, data within this kinematical limit were obtained by the HERA collaborations H1 and ZEUS. Based on the results of $T_{00}$ and $T_{11}$ of ref.~\cite{Besse:2013muy}, we can describe the polarized cross-sections and spin density matrix elements, which we can compare with H1 and ZEUS data.
 
\begin{figure}[htbp]
	\centering
\psfrag{P}[cc][cc]{$p$}
\psfrag{P'}[cc][cc]{$p'$}
\psfrag{Gam}[cc][cc]{$\gamma^*$}
\psfrag{Rho}[cc][cc]{$\rho$}
\psfrag{k}[cc][cc]{$k_1$}
\psfrag{k2}[cc][cc]{$k_2$}
\psfrag{Phi}[cc][cc]{${\Phi^{\gamma^*_{\lambda_{\gamma}}\to\rho_{\lambda_{\rho}}}}$}
\psfrag{Phibis}[cc][cc]{$\mathcal{F}(x,\kb)$}
\includegraphics[width=0.5\linewidth]{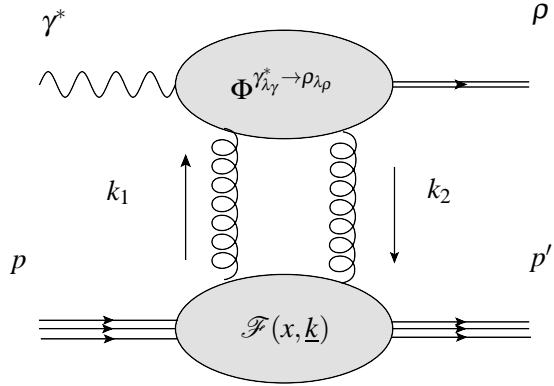}
\caption{Impact factor representation of the helicity amplitudes.}
	\label{ImpFacto}
\end{figure}

  Our approach is based on the impact factor representation of the helicity amplitudes
\begin{align}
\label{kT}
T_{\lambda_{\rho}\lambda_{\gamma}}= is \int \frac{d^2\kb}{(\kb^2)^2} \, \Phi^{\gamma^*_{\lambda_{\gamma}}\to\rho_{\lambda_{\rho}}}(\kb) \, \mathcal{F}(x,\kb)\,,
\end{align}
where $\mathcal{F}(x,\kb)$ stands for the unintegrated gluon density\footnote{We denote $\kb$ the euclidean transverse vector such that $\kb^2=-k_{\perp}^2$.} and where 
\begin{align*}
\Phi^{\gamma^*_{\lambda_{\gamma}}\to\rho_{\lambda_{\rho}}}=\frac{1}{2s} \int \frac{d \kappa}{2\pi} i \mathcal{M}(\gamma^*(\lambda_{\gamma},q)+g(k_1)\to \rho(\lambda_{\rho},p_1)+g(k_2))\,,
\end{align*}
is the $\gamma^*(\lambda_{\gamma})\to\rho(\lambda_{\rho})$ impact factor associated to the matrix element $\mathcal{M}$ as illustrated in fig. \ref{ImpFacto}.

The hard virtuality of the photon justifies a perturbative treatment of the impact factor $\Phi^{\gamma^*\to\rho}$ using light-cone collinear factorization (LCCF). The large distance non-perturbative physics of hadronization of the partons into a $\rho$ meson is  encoded by the $\rho$ meson distribution amplitudes. For the production of a longitudinally polarized $\rho$ meson, the calculations of the impact factors $\Phi^{\gamma^*_{L}\to \rho_L}$ and $\Phi^{\gamma^*_T\to \rho_L}$ were performed in ref.~\cite{GinzburgPanfilSerbo}. At leading twist, the production of a longitudinally polarized $\rho$ meson involves the twist 2 distribution amplitude, while the next term of the twist expansion is only of twist 4. The production of a transversely polarized $\rho$ meson is particularly interesting as the leading twist non-vanishing contribution is of twist 3. It requires more involved calculations and the impact factor $\Phi^{\gamma^*_T\to\rho_T}$ was calculated in refs.~\cite{Anikin2009, Anikin2010} in the limit $t\sim 0$. The LCCF approach allows to get a gauge invariant treatment of the impact factors as it was checked in refs.~\cite{Anikin2009, Anikin2010}. The relevant twist 2 and twist 3 non-local operators on the light-cone are obtained from the Taylor expansion of the hard sub-processes around the light-cone direction of the $\rho$ meson. The hard sub-processes involve quark-antiquark and quark-antiquark-gluon intermediate Fock states up to twist 3. 

A first model was made in ref.~\cite{Anikin2011} using these impact factors in combination with the model of distribution amplitudes of ref.~\cite{Ball:1998sk} and a model for the proton impact factor inspired from ref.~\cite{GunionSoper}. The results of this model underlined the effect of soft gluon exchange in $t-$channel. The sizable contribution of the exchange of soft gluons ($\kb^2\lesssim 1\;$GeV$^2$) indicates that we should include saturation effects in this first approach. It is convenient, in order to include saturation effects, to work in the impact parameter space (i.e. transverse coordinate space) in order to express the amplitude in terms of color dipole degrees of freedom as it is done in the color dipole picture. The saturation effects are then included in the dipole scattering amplitude with the nucleon target, and amount in suppressing large dipole size contributions.

We have shown in ref.~\cite{Besse2012} that the impact factors $\Phi^{\gamma^*_{L}\to \rho_L}$ and $\Phi^{\gamma^*_T\to \rho_T}$ in the LCCF frame and in the limit $t\sim 0$, results in the convolutions
\begin{eqnarray}
\label{phiLpsi}
\Phi^{\gamma^{*}_L \rightarrow \rho_L}(\kb,Q,\mu^2)&=&\left(\frac{\delta^{ab}}{2}\right)\int dy \int d \rb\,\, \psi^{\gamma^*_L\to\rho_L}_{(q\qb)}(y,\rb;Q,\mu^2)\, \mathcal{A}(\rb,\kb)\,,\\
\Phi^{\gamma^{*}_T \rightarrow \rho_T}(\kb,Q,\mu^2)&=&\left(\frac{\delta^{ab}}{2}\right)\int dy \int d \rb\,\, \psi^{\gamma^*_T\to\rho_T}_{(q\qb)}(y,\rb;Q,\mu^2) \,\mathcal{A}(\rb,\kb)\nn\\
&&\hspace{-2cm}+\left(\frac{\delta^{ab}}{2}\right)\int dy_2 \int dy_1 \int d \rb \,\, \psi^{\gamma^*_T\to\rho_T}_{(q\qb g)}(y_1,y_2,\rb;Q,\mu^2) \mathcal{A}(\rb,\kb)\,,
\label{phiTpsi}
\end{eqnarray}
once they are expressed 
in the impact parameter representation. In eqs.~(\ref{phiLpsi}, \ref{phiTpsi}), $\mathcal{A}(\rb,\kb)$ is the dipole scattering amplitude with the two gluons of the $t-$channel while the functions 

$\psi^{\gamma^*_L\to\rho_L}_{(q\qb)}(y,\rb;Q,\mu^2)$, $\psi^{\gamma^*_T\to\rho_T}_{(q\qb)}(y,\rb;Q,\mu^2)$ and $\psi^{\gamma^*_T\to\rho_T}_{(q\qb g)}(y_1,y_2,\rb;Q,\mu^2)$ are respectively the amplitude of probability to have an interacting dipole of size $\rb$ constituted from the quark-antiquark or quark-antiquark-gluon intermediate states. These amplitudes $\psi^{\gamma^*_{L,T}\to\rho_{L,T}}_{(q\qb (g))}$ are the results of the LCCF approach, they depend on the distribution amplitudes of the $\rho$ meson. It was checked that the results eqs.~(\ref{phiLpsi}, \ref{phiTpsi}) of ref.~\cite{Besse2012} are equivalent to the results found in refs.~\cite{Anikin2009, Anikin2010}. 
Combining eqs.~(\ref{kT}--\ref{phiTpsi}) leads to the following expressions for the helicity amplitudes
\begin{eqnarray}
\label{T00Lpsi}
T_{00}\!&\!=\!&\!s\!\int dy \int d \rb\, \psi^{\gamma^*_L\to\rho_L}_{(q\qb)}(y,\rb;Q,\mu^2) \,\hat{\sigma}(x,\rb)\,,\\
\label{T11Tpsi}T_{11}\!&\!=s\!&\! \int d \rb\left[\int \! dy\, \psi^{\gamma^*_T\to\rho_T}_{(q\qb)}(y,\rb;Q,\mu^2)
\right.\\&&\hspace{-1cm}
\left.+\int \! dy_2 \int \! dy_1  \psi^{\gamma^*_T\to\rho_T}_{(q\qb g)}(y_1,y_2,\rb;Q,\mu^2)\right] \hat{\sigma}(x,\rb)\,,\nn
\end{eqnarray}
where $\hat{\sigma}(x,\rb)$ is the dipole cross-section. This dipole cross-section is a universal quantity which is fitted in the forward limit on inclusive observables. 

Note that we neglect here skewness effects in order to identify the dipole scattering amplitude appearing in our treatment of the exclusive production of $\rho$ meson with the dipole cross-section in the forward limit appearing in deep inelastic scattering cross-sections. 

We can access then to polarized cross-sections
\begin{eqnarray*}
\sigma_L&=&\frac{1}{b(Q^2)}\frac{\left| T_{00}\right|^2}{16 \pi s^2}\,,\\
\sigma_T&=&\frac{1}{b(Q^2)}\frac{\left|T_{11}\right|^2}{16 \pi s^2}\,,
\end{eqnarray*}
where $b(Q^2)$ is the $b-$slope determined experimentally from the $t-$dependence of the differential cross-section. We used the $b-$slope measured by H1 \cite{H1}.

\section{Results}
\begin{figure}[h]
	\centering
	\begin{tabular}{c}
		\includegraphics[width=0.7\textwidth]{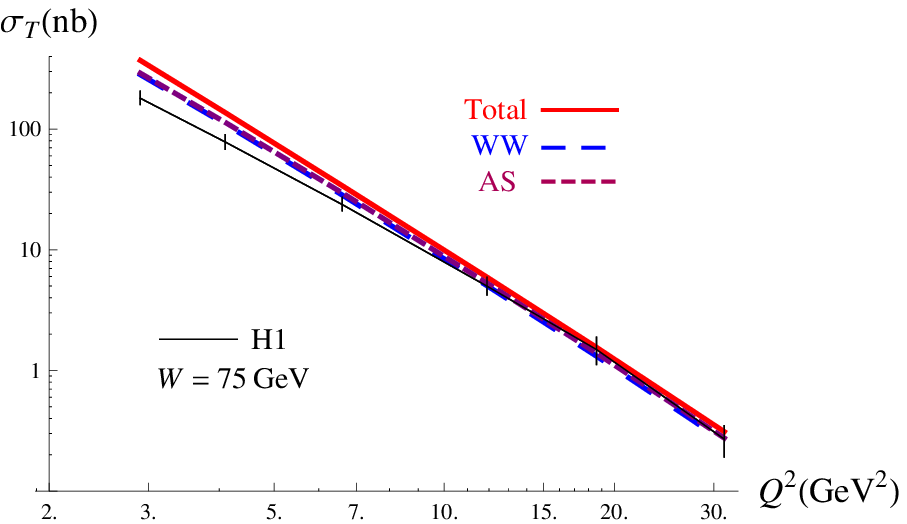}\\
		\includegraphics[width=0.7\textwidth]{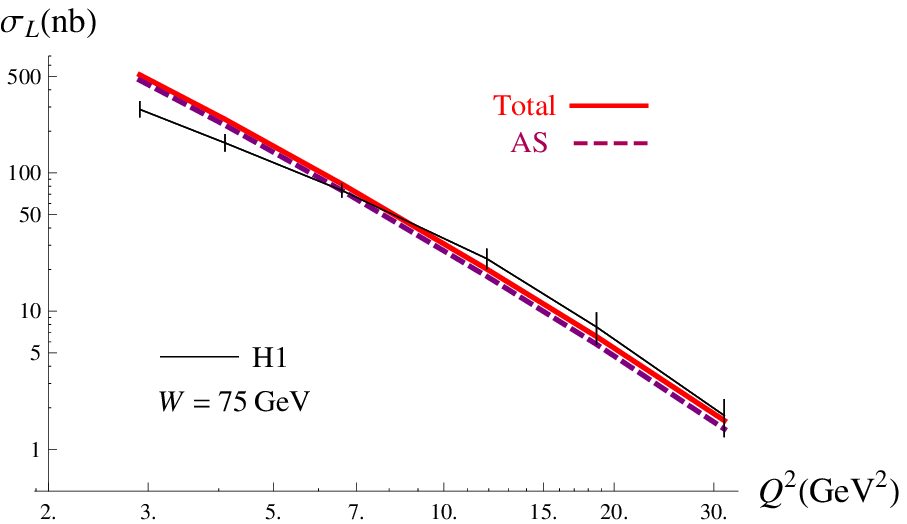}
	\end{tabular}
	\caption{Left: Total, WW and AS contributions to $\sigma_T$ vs $Q^2$, compared to H1 \cite{H1} data. Right: Total and AS twist~2 contributions to $\sigma_L$ vs $Q^2$ compared to H1 data. }	\label{Fig3}
\end{figure}

In fig.~\ref{Fig3} are shown the predictions we obtained in ref.~\cite{Besse:2013muy} compared to H1 data \cite{H1}. The asymptotic cross-section (AS) corresponds to the case where the renormalization scale in the model of distribution amplitudes is chosen asymptotically large. The full twist 3 (Total) result for $\sigma_T$ and the twist 2 result (Total) for $\sigma_L$ are the predictions with the choice of renormalization scale $\mu^2=(Q^2+m^2_{\rho})/4$. Finally, the WW result for the cross-section $\sigma_T$ corresponds to the contribution from the distribution amplitudes in the Wandzura-Wilczek approximation, which consist in neglecting contributions from the quark-antiquark-gluon distribution amplitudes. 

We used for these predictions a model of dipole cross-section \cite{Albacete:2010sy} with a very good $\chi^2/dof\sim 1.2 $ fitted on inclusive and longitudinal structure functions of deep inelastic scattering as well as the model of distribution amplitudes of Ball, Braun, Koike and Tanaka~\cite{Ball:1998sk}. 

\begin{figure}[h]
	\centering
	\begin{tabular}{c}
		\includegraphics[width=0.7\textwidth]{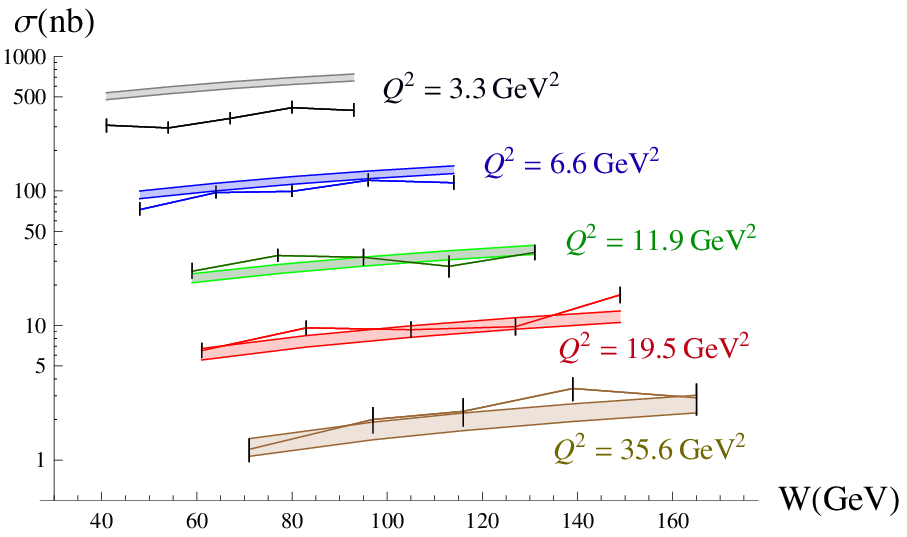}	\\
	\includegraphics[width=0.7\textwidth]{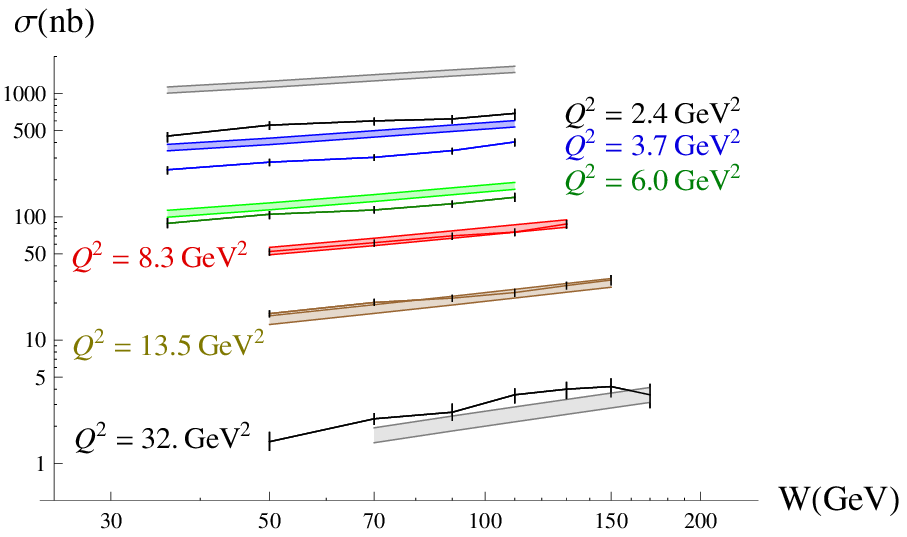}
	\end{tabular}
	\caption{
	Predictions for the total cross-section $\sigma$ vs $W$ compared to H1 \cite{H1} (top) and ZEUS \cite{ZEUS} (bottom) data.}
	\label{Fig4}
\end{figure}

We see that results are in good agreement with the data of H1 for virtualities larger than $\sim 5\;$GeV$^2$. The $W$ dependence of the total cross-section is also well reproduced as shown in fig.~\ref{Fig4}, as long as $Q^2\gtrsim 5\;$GeV$^2$.

\section{Conclusions}
Our presented model is based on LCCF calculations of the impact factors $\Phi^{\gamma^*_{\lambda_{\gamma}}\to\rho_{\lambda_{\rho}}}$ in the impact parameter representation. It leads to very good predictions for a large range in $Q^2$, with no free parameters. The discrepancy for low $Q^2$ could be due to higher twist corrections. A generalization of the approach to other vector meson production such as the $\phi$ meson could be done in the same way by taking the right set of distribution amplitudes. This work can be generalized also for the exchange of a transverse momentum exchange in $t-$channel. This would allow to link amplitudes of probability to get a dipole of size $\rb$ with a given impact parameter $\underline{b}$ from the target, in combination with dipole models that have a $\underline{b}$ dependence. The additional impact parameter $\underline{b}$ degree of freedom allows to get information on the transverse distribution of gluons in the target.

\section{Acknowledgment}
We thank B. Duclou\'e, K. Golec-Biernat, C. Marquet, S. Munier and B. Pire for interesting discussions and comments on this work. This work is supported by the P2IO consortium, the Polish Grant NCN No DEC-2011/01/B/ST2/03915, the Joint Research Activity Study of Strongly Interacting Matter (acronym HadronPhysics3, Grant 283286) under the Seventh Framework Programme of the European Community and the French grant ANR PARTONS (ANR-12-MONU-0008-01).

\end{document}